\documentclass[graybox]{svmult}

\usepackage{mathptmx}       
\usepackage{helvet}         
\usepackage{courier}        
\usepackage{type1cm}        
\usepackage{makeidx}         
\usepackage{graphicx}        
\usepackage{multicol}        
\usepackage[bottom]{footmisc}

\usepackage[numbers,sort&compress,sectionbib]{natbib}

\makeindex             

\usepackage{bm} 
\usepackage{amsmath}
\usepackage{amsfonts}
\usepackage{amssymb}
\usepackage{dsfont}
\usepackage{subcaption}
\usepackage{booktabs}
\usepackage{graphicx}
\usepackage{multirow}

\newcommand{\vv}{\mathbf{v}}
\newcommand{\x}{\mathbf{x}}
\newcommand{\rr}{\mathbf{r}}

\usepackage{url}

\graphicspath{{./figures/}}

\begin{document}

\title*{Learning representations of molecules and materials with atomistic neural networks}
\author{Kristof T. Sch\"utt \and Alexandre Tkatchenko \and Klaus-Robert M\"uller}
\institute{Kristof T. Sch\"utt \at Machine Learning Group, Technische Universit\"at Berlin, 10587 Berlin, Germany, \email{kristof.schuett@tu-berlin.de}
\and Alexandre Tkatchenko \at Physics and Materials Science Research Unit, University of Luxembourg, L-1511 Luxembourg, Luxembourg \email{alexandre.tkatchenko@uni.lu} \and Klaus-Robert M\"uller \at Machine Learning Group, Technische Universit\"at Berlin, 10587 Berlin, Germany \\
Max-Planck-Institut f\"ur Informatik, Saarbr\"ucken, Germany \\
Department of Brain and Cognitive Engineering, Korea University, Anam-dong, Seongbuk-gu, Seoul 02841, Korea
\email{klaus-robert.mueller@tu-berlin.de}}

%
%
\maketitle

\abstract*{Deep Learning has been shown to learn efficient representations for structured data such as image, text or audio.
	In this chapter, we present neural network architectures that are able to learn efficient representations of molecules and materials.
	In particular, the continuous-filter convolutional network SchNet accurately predicts chemical properties across compositional and configurational space on a variety of datasets.
	Beyond that, we analyze the obtained representations to find evidence that their spatial and chemical properties agree with chemical intuition.}

\abstract{Deep Learning has been shown to learn efficient representations for structured data such as image, text or audio.
	In this chapter, we present neural network architectures that are able to learn efficient representations of molecules and materials.
	In particular, the continuous-filter convolutional network SchNet accurately predicts chemical properties across compositional and configurational space on a variety of datasets.
	Beyond that, we analyze the obtained representations to find evidence that their spatial and chemical properties agree with chemical intuition.}

\section{Introduction}\label{sec:atomistic}
In recent years, machine learning has been successfully applied to the prediction of chemical properties for molecules and materials~\cite{schutt2014represent,huo2017unified,faber2018alchemical,de2016comparing,morawietz2016van,gastegger2017machine,faber2017prediction,podryabinkin2017active,brockherde2017bypassing,bartok2017machine,schutt2018schnet,chmiela2018towards,ziletti2018insightful,dragoni2018achieving}.
A significant part of the research has been dedicated to engineering features that characterize global molecular similarity~\cite{bartok2010gaussian,rupp2012fast,montavon2012learning,hansen2013assessment,Hansen-JCPL,chmiela2017machine} or local chemical environments~\cite{behler2007generalized,bartok2013representing,gastegger2018wacsf} based on atomic positions.
Then, a non-linear regression method -- such as kernel ridge regression or a neural network -- is used to correlate these features with the chemical property of interest.

A common approach to model atomistic systems is to decompose them into local environments, where a chemical property is expressed by a partitioning into latent atom-wise contributions.
Based on these contributions, the original property is then reconstructed via a physically motivated aggregation layer.
E.g., Behler-Parinello networks~\cite{behler2007generalized} or the SOAP kernel~\cite{bartok2013representing} decompose the total energy in terms of atomistic contributions
\begin{align}
E &= \sum_{i=1}^{n_\text{atoms}} E_i \label{eq:localcontributions}.
\end{align}
Atomic forces can be directly obtained as negative derivatives of the energy model.
While this is often a suitable partitioning of extensive properties, intensive or per-atom properties can be modeled as the mean
\begin{align}
P &= \frac{1}{n_\text{atoms}} \sum_{i=1}^{n_\text{atoms}} P_i \label{eq:localcontributions2}.
\end{align}
However, this might still not be a sufficient solution for global molecular properties such as HOMO-LUMO gaps or excitation energies~\cite{pronobis2018capturing}.
To obtain a better performance, output models that incorporate property-specific prior knowledge should be used.
E.g., the dipole moment can be written as 
\begin{equation}
\bm{\mu} = \sum^N_i q_i \mathbf{r}_i. \label{eq:atmu}
\end{equation}
such that the atomistic neural network needs to predict atomic charges $q_i$~\cite{gastegger2017machine,sifain2018discovering,yao2018tensormol,schutt2018quantum}.

The various atomistic models differ in how they obtain the energy contributions $E_i$, usually employing manually crafted atomistic descriptors.
In contrast to such descriptor-based approaches, this chapter focuses on atomistic neural network architectures that learn efficient representations of molecules and materials \emph{end-to-end} -- i.e., directly from atom types $Z_i$ and positions $\rr_i$ -- while delivering accurate predictions of chemical properties across compositional and configurational space~\cite{schutt2017quantum,schutt2017schnet,schutt2018schnet}.
The presented models will encode important invariances, e.g. towards rotation and translation, directly into the deep learning architecture and obtain predicted property from physically motivated output layers.
Finally, we will obtain spatially and chemically resolved insights from the learned representations regarding the inner workings of the neural network as well as the underlying data.

\section{The deep tensor neural network framework}

In order to construct atom-centered representations $\x_i \in \mathbb{R}^F$, where $i$ is the index of the center atom and $F$ the number of feature dimension, a straight-forward approach is to expand the atomistic environment in terms of n-body interactions~\cite{pukrittayakamee2009simultaneous,malshe2009development}, which can be written in general as
\begin{equation}
\x_i = f^{(1)}(Z_i) + \sum_{j \neq i} f^{(2)}((Z_i, \rr_i), (Z_j, \rr_j)) + \sum_{\substack{j,k \neq i \\ k \neq j}} f^{(3)}((Z_i, \rr_i), (Z_j, \rr_j), (Z_k, \rr_k)) + \dots \label{eq:mbe}
\end{equation}
However, such an approach requires to define explicit n-body models $f^{(n)}$ (e.g. using neural networks) as well as computing of a large number of higher-order terms of the atom coordinates.
At the same time, all many-body networks must respect the invariances w.r.t. rotation, translation and the permutation of atoms.

An alternative approach is to incorporate higher-order interactions in a recursive fashion.
Instead of explicitly modeling an $n$-body neural network, we design an interaction network $\vv: \mathbb{R}^{F} \times \mathbb{R} \rightarrow \mathbb{R}^{F}$ that we use to model perturbations
\begin{align}\label{eq:dtnn:perturbation}
\x_i^{(t+1)} &= \x_i^{(t)} + \vv^{(t)}(\x_1^{(t)}, \rr_{i1}, \dots, \x_n^{(t)}, \rr_{in}),
\end{align}
of the chemical environment $\x_i^{(t)}$ by its neighboring environments $\x_j^{(t)}$ depending on their relative position $\rr_{ij} = \rr_j - \rr_i$.
On this basis, we define the \emph{deep tensor neural network} (DTNN) framework~\cite{schutt2017quantum}:
\begin{enumerate}
	\item Use an embedding depending on the type of the center atom
	\[
	\mathbf{x}_i^{(0)} = \mathbf{A}_{Z_i} \in \mathbb{R}^d
	\]
	for the initial representation of local chemical environment $i$.
	This corresponds to the 1-body terms in Eq.~\ref{eq:mbe}.
	\item Refine the embeddings repeatedly using the interaction networks from Eq.~\ref{eq:dtnn:perturbation}
	\item Arriving at the final embedding $\x_i^T$ after $T$ interaction refinements, predict the desired chemical property using a property-specific output network (as described in Section~\ref{sec:atomistic}).
\end{enumerate}
The embedding matrix $A$ as well as the parameters of the interaction networks $\vv^{(t)}$ and the output network are learned during the training procedure.
This framework allows for a family of atomistic neural network models -- such as the Deep Tensor Neural Network~\cite{schutt2017quantum} and SchNet~\cite{schutt2017schnet,schutt2018schnet} -- that differ in how the interactions $\vv^{(t)}$ are modeled and the predictions are obtained from the atomistic representations $\x_i^T$.

\section{SchNet}
\begin{figure}
	\centering
	\includegraphics[width=0.8\textwidth]{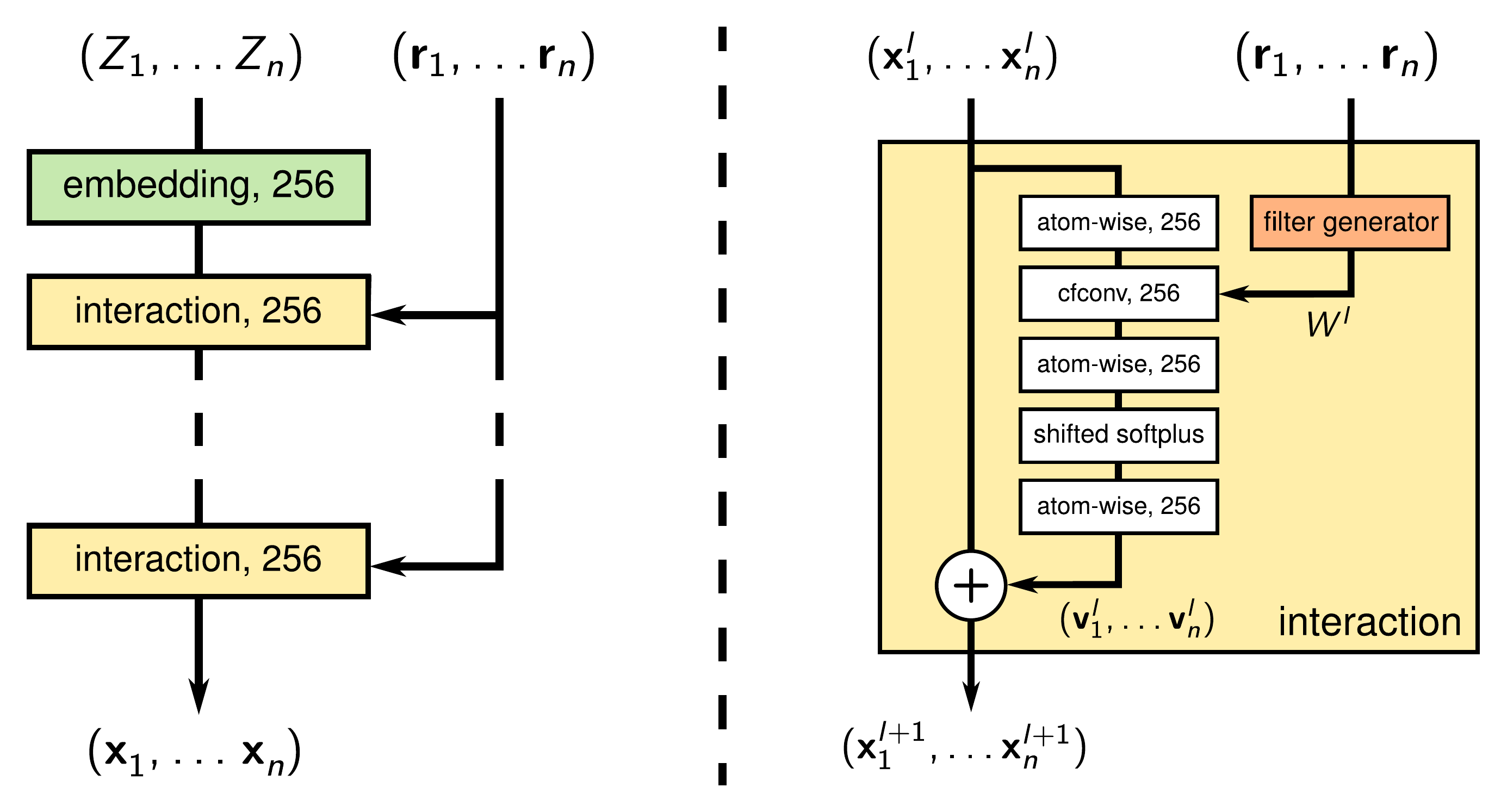}
	\caption{The illustration shows an architectural overview of SchNet (left), the interaction block (right). The shifted softplus activation function is defined as $\text{ssp}(x) = \ln(0.5e^x + 0.5)$. The number of neurons used in the employed SchNet models is given for each parameterized layer.}\label{fig:schnet:architecture}
\end{figure}

Building upon the principles of the previously described DTNN framework, we propose SchNet as a convolutional neural network architecture for learning representations for molecules and materials.
Fig.~\ref{fig:schnet:architecture} depicts an overview of the SchNet architecture as well as how the interaction refinements are modeled by interaction blocks shown on the right.
In the following, we will introduce the main component of SchNet -- the continuous-filter convolutional layer -- before describing how these are used to construct the interaction blocks.

\subsection{Continuous-filter convolutional layers}

The commonly used convolutional layers~\cite{lecun1989backpropagation} employ discrete filter tensors since they are usually applied to data that is sampled on a grid, such as digital images, video or audio.
However, such layers are not applicable for atomistic systems, since atoms can be located at arbitrary positions in space.
E.g. when predicting the potential energy, the output of a convolutional layer might change rapidly when an atom moves from one grid cell to the next.
Especially when we aim to predict a smooth potential energy surface, a continuous and differentiable representation is required.
For this reason, we use a convolutional layer employing a continuous filter function in order to model the interactions.

Given the representations $\x^l_i$ of the chemical environment of atom $i$ at position $\rr_i$ and layer $l$ of the neural network, the atomistic system can be described by a function
\begin{equation}
\rho^l(\rr) = \sum_{i=1}^{n_\text{atoms}} \delta(\rr - \rr_i) \x^l_i.\label{eq:rhodef}
\end{equation}

In order to include the interactions of the atom-centered environments, we convolve $\rho: \mathbb{R}^3 \rightarrow \mathbb{R}^F$ and a spatial filter $W: \mathbb{R}^3 \rightarrow \mathbb{R}^F$ as element-wise
\begin{equation}
(\rho * W)(\rr) = \int\limits_{\rr_a \in \mathbb{R}^3} \rho(\rr_a) \circ W(\rr-\rr_a) d\rr_a, \label{eq:contconv}
\end{equation}
where "$\circ$" is the element-wise product.
Here, the filter function $W$ describes the interaction of a representation $\x_i$ with an atom at the relative position $\rr - \rr_i$.
The filter functions can be modeled by a \emph{filter-generating} neural network similar to those used in dynamic filter networks~\cite{BrabandereJTG16}.
Considering the discrete location of atoms in Eq.~\ref{eq:rhodef}, we obtain
\begin{align}
(\rho^l * W)(\rr)
&= \sum_{j=1}^{n_\text{atoms}} \int\limits_{\rr_a \in \mathbb{R}^3}  \delta(\rr_a - \rr_j) \x^l_j \circ W(\rr-\rr_a) d\rr_a  \notag \\
&= \sum_{j=1}^{n_\text{atoms}} \x^l_j \circ W(\rr-\rr_j) \label{eq:contconvatom}
\end{align}
This yields a function representing how the atoms of the system act on another location in space.
To obtain the rotationally-invariant interactions between atoms, we 
\begin{equation}
\x_i^{l+1} = (\rho^l * W^l)(\rr_i) = \sum_{j=1}^{n_\text{atoms}} \x^l_j \circ W(\rr_j - \rr_i), \label{eq:atomconv}
\end{equation}
i.e., we evaluate the convolution at discrete locations in space using continuous, radial filters.

\subsection{Interaction blocks}\label{sec:schnet:interactions}
After introducing continuous-filter convolutional layers, we go on to construct the interaction blocks.
Besides convolutions, we employ atom-wise, fully-connected layers
\begin{equation}
\x^{(l+1)}_i = W^{(l)} \x^{(l)}_i + \mathbf{b}^{(l)}
\end{equation}
that are applied separately to each atom $i$ with tied weights $W^{(l)}$.
\begin{figure}[tb]
	\centering
	\includegraphics[width=\textwidth]{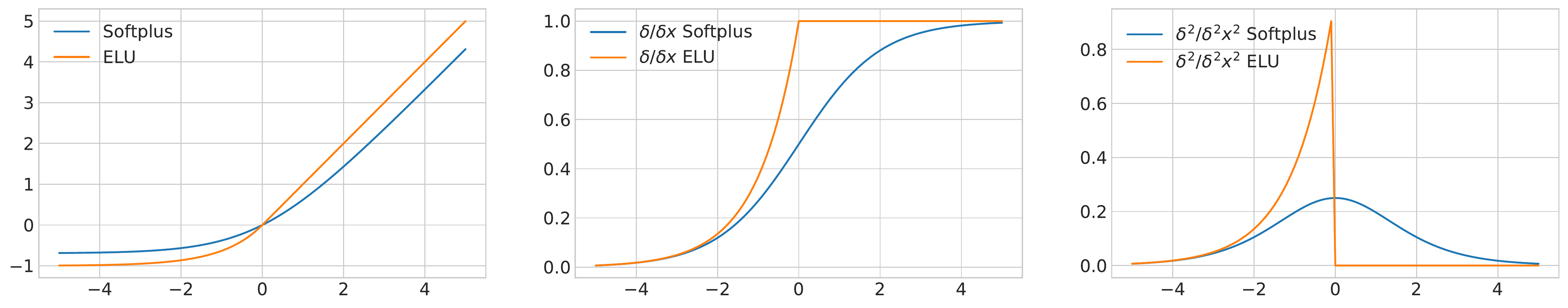}
	\caption{\textbf{Comparison of shifted softplus and ELU activation function.} We show plots of the activation functions (left), and their first (middle) and second derivatives (right).}\label{fig:schnet:activation}
\end{figure}
Throughout the network, we use softplus non-linearities~\cite{dugas2001incorporating} that are shifted
\begin{equation}
f(x) = \ln \left(\frac{1}{2} e^x + \frac{1}{2} \right)
\end{equation}
in order to conserve zero-activations: $f(0)=0$.
Fig.~\ref{fig:schnet:activation} shows this activation function compared to exponential linear units (ELU)~\cite{clevert2015fast}:
\begin{equation}
f(x) = \begin{cases}
e^x-1 & \text{if } x<0 \\
x & \text{otherwise}
\end{cases}
\end{equation} 
The derivatives for ELU and softplus are shown in the middle and right panel of Fig.~\ref{fig:schnet:activation}, respectively.
A crucial difference is that the softplus are smooth while ELUs exhibit only first-order continuity.
However, the higher-order differentiability of the model, and therefore also of the employed activation functions, is crucial for the prediction of atomic forces or vibrational frequencies.

Fig.~\ref{fig:schnet:architecture} (right) shows how the interaction block is assembled from these components.
Since the continuous-filter convolutional layers are applied feature-wise, we achieve the mixing of feature maps by atom-wise layers before and after the convolution.
This is analogous to depth-wise separable convolutional layers in Xception nets~\cite{chollet2017xception} which could outperform the architecturally similar InceptionV3~\cite{szegedy2016rethinking} on the ImageNet dataset~\cite{deng2009imagenet} while having less parameters.
Most importantly, feature-wise convolutional layers reduce the number of filters, which significantly reduces the computational cost.
This is particularly important for continuous-filter convolutions, where each filter has to be computed by a filter-generating network.

\subsection{Filter-generating networks}\label{sec:schnet:filters}
The architecture of the filter-generating network significantly influences the properties of the predicted filters and, consequently, the learned atomic interactions.
Therefore, we can incorporate invariances or prior chemical knowledge into the filter.
In the following, we describe the considerations that went into designing the SchNet filter-generating networks.

\subsubsection{Self-interaction}
In an interatomic potential, we aim to avoid self-interaction of atoms, since this is fundamentally different than the interaction with other atoms.
We can encode this in the filter network by constraining the filter-network such that $W(\rr_i-\rr_j) = 0$ for $\rr_i = \rr_j$.
Since two distinct atoms can not be at the same position, this is a unambiguous condition to exclude self-interaction.
This is equivalent to modifying Eq.~\ref{eq:atomconv} to exclude the center atom of the environment from the sum:
\begin{equation}
\x_i^{l+1} = \sum_{j \neq i} \x^l_j \circ W(\rr_j - \rr_i), \label{eq:atomconv2}
\end{equation}

\subsubsection{Rotational invariance}
As the input to the filter $W: \mathbb{R}^3 \rightarrow \mathbb{R}$ is only invariant to translations of the molecule, we additionally need to consider rotational invariance.
We achieve this by using interatomic distances $r_{ij}$ as input to the filter network, resulting in radial filters $W: \mathbb{R} \rightarrow \mathbb{R}^F$.

\subsubsection{Local distance regimes}

\begin{figure}[tb]
	\centering
	\includegraphics[width=0.65\textwidth]{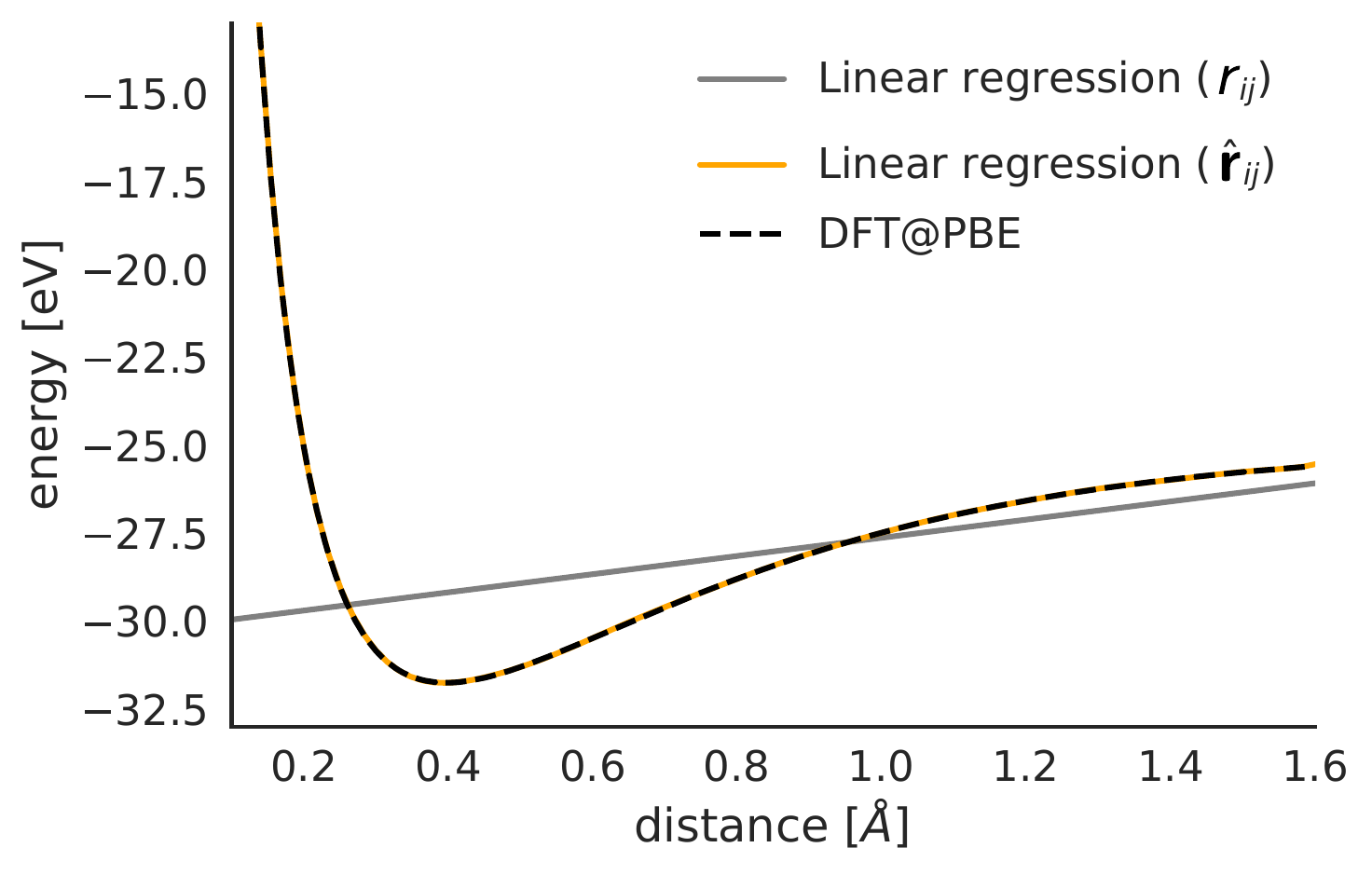}
	\caption{Comparison of features for regression of bond stretching energies of H$_2$.
		We use scalar distances $r_{ij}$ and distances in a radial basis $\hat{\rr}_{ij}$ with $\Delta\mu = 0.1$ and $\gamma=10$ as features, respectively.
		The energies were computed by Brockherde et al. \cite{brockherde2017bypassing} with DFT at the PBE level of theory.
		\label{fig:bondstretching}
	}
\end{figure}

In the spirit of radial basis function (RBF) networks~\cite{broomhead1988multivariable,moody1989fast}, the filter-generating neural network $W(r_{ij})$ first expands the pair-wise distances
\begin{equation}
\hat{\mathbf{r}}_{ij} = \left[ \exp ( -\gamma (r_{ij} - k \Delta\mu )^2 ) \right]_{0 \leq k \leq r_{\text{cut}}/\Delta \mu}, \label{eq:radialbasis}
\end{equation}
with $\Delta \mu$ being the spacing of Gaussians with scale $\gamma$ on a grid ranging from 0 to the distance cutoff $r_\text{cut}$.
This helps to decouple the various regimes of atomic interactions and allow for an easier starting point for the training procedure.
On top of the RBF expansion, we apply two fully-connected layers with softplus activation functions.

As an illustrative example, Fig.~\ref{fig:bondstretching} shows two linear models fitted to the potential energy surface of H$_2$.
Using the distance as feature directly, we obviously capture only a linear relationship. 
However, the expanded RBF feature space allows us to obtain a smooth and accurate fit of the potential.

From an alternative viewpoint, if we initialize a neural network with the usual weight distributions and non-linearities, the resulting function is almost linear before training as the neuron activations are close to zero.
Therefore, the filter values would be strongly correlated, leading to a plateauing cost function at the beginning of training.
Radial basis functions solve this issue by decoupling the various distance regimes.

\subsubsection{Cutoffs}

While in principle the size of the filter in a continuous-filter convolutional layer can be infinite, there are natural limitations on how such a filter can be trained.
The interatomic distances in a dataset of molecules are upper-bound by the size of the largest molecule.
More importantly, we can not consider interactions with an infinite amount of atoms in case of atomistic systems with periodic boundary conditions.
Therefore, it is often beneficial or even required to restrict the filter size using a distance cutoff.

While it is certainly possible to apply a hard cutoff, this may lead to rapidly changing energies in molecular dynamics simulations.
Therefore, we apply a cosine cutoff function to the filter, to obtain a local filter
\begin{align}
f_\text{cut}(r_{ij}) &= \frac{1}{2} \cos \left( \frac{r_{ij}}{r_\text{cut}} \pi \right) + \frac{1}{2} \\
W_\text{local}(r_{ij}) &=   W(r_{ij}) f_\text{cut}(r_{ij})
\end{align}

\subsubsection{Periodic boundary conditions (PBC)}
For materials, we have to respect the PBCs when convolving with the interactions, i.e. we have to include interactions with periodic sites of neighboring unit cells:
Due to the linearity of the convolution, we can move the sum over periodic images into the filter.
Given atomistic representations $\x_i = \x_{ia} = \x_{ib}$ of site $i$ for unit cells $a$ and $b$, we obtain
\begin{align}
\x^{l+1}_{i} = \x^{l+1}_{im} &= \sum_{j = 0}^{n_\text{atoms}}  \sum_{b = 0}^{n_\text{cells}} \x_{jb}^l \circ \tilde{W}^l(\rr_{jb} - \rr_{ia}) \notag \\
&= \sum_{j = 0}^{n_\text{atoms}} \x^l_j \circ \underbrace{\left( \sum_{b = 0}^{n_\text{cells}} \tilde{W}^l(\rr_{jb} - \rr_{ia}) \right)}_{W}.
\end{align}
When using a hard cutoff, we have found that the filter needs to be normalized with respect to the number of neighboring atoms $n_\text{nbh}$ for the training to converge:
\begin{align}
W_\text{normalized}(r_{ij}) = \frac{1}{n_\text{nbh}}W(r_{ij})
\end{align}
However, this is not necessary, when using a cosine cutoff function, as shown above.

\section{Analysis of the representation}
Having introduced the SchNet architecture, we go on to analyze the representations that have been learned by training the models on QM9 -- a dataset of ~130k small organic molecules with up to nine heavy atoms~\cite{ramakrishnan2014quantum} -- as well as a molecular dynamics trajectory of aspirin~\cite{chmiela2017machine}.
If not given otherwise, we use six interaction blocks and atomistic representations $\x_i \in \mathbb{R}^{256}$.
The models have been trained using stochastic gradient descent with warm restarts~\cite{loshchilov2016sgdr} and the ADAM optimizer~\cite{kingma2014adam}.

\subsection{Locality of the representation}
As described above, atomistic models decompose the representation into local chemical environments.
Since SchNet is able to learn a representation of such an environment, the locality of the representation may depend on whether a cutoff was used as well as the training data.

\begin{table}[tb]
	\caption{Mean absolute (MAE) and root mean squared errors (RMSE) of SchNet with and without cosine cutoff for various datasets over three repetitions. For the Materials Project data, we use a smaller model ($\x_i \in \mathbb{R}^{64}$) and compare the cosine cutoff to a normalized filter with hard cutoff. We give the number of used reference calculations N, i.e. the size of the combined training and validation set.\label{tab:results}}
	\setlength{\tabcolsep}{.4em}
	\begin{tabular}{lllrrr}
		\toprule
		\textbf{Dataset} & \textbf{Property} & \textbf{Unit} & \textbf{$r_\text{cut}$ [{\AA}]} & \hspace{.8em} \textbf{MAE} & \hspace{.8em} \textbf{RMSE} \\ \midrule
		\multirow{4}{*}{QM9 (N=110k)} & \multirow{2}{*}{U$_0$} & \multirow{2}{*}{kcal mol$^{-1}$} & -- & 0.259 & 0.599 \\ 
		&  &  & $5$ & \textbf{0.218} & \textbf{0.518} \\ \cmidrule{2-6}
		& \multirow{2}{*}{$\mu$} & \multirow{2}{*}{Debye} & -- & 0.019 & 0.037 \\
		&  &  & $5$ & \textbf{0.017} & \textbf{0.033} \\
		\midrule
		\multirow{4}{*}{Aspirin (N=1k)} & \multirow{2}{*}{total energy} & \multirow{2}{*}{kcal mol$^{-1}$} & -- & 0.438 & 0.592 \\
		&  &  & 5 & \textbf{0.402} & \textbf{0.537} \\ \cmidrule{2-6}
		& \multirow{2}{*}{atomic forces} & \multirow{2}{*}{kcal mol$^{-1}${\AA}$^{-1}$} & -- & 1.359 & 1.929 \\
		&  &  & 5 & \textbf{0.916} & \textbf{1.356} \\	\midrule
		\multirow{4}{*}{Aspirin (N=50k)} & \multirow{2}{*}{total energy} & \multirow{2}{*}{kcal mol$^{-1}$} & -- & \textbf{0.088} & \textbf{0.113} \\
		&  &  & 5 & 0.102 & 0.130 \\ \cmidrule{2-6}
		& \multirow{2}{*}{atomic forces} & \multirow{2}{*}{kcal mol$^{-1}${\AA}$^{-1}$} & -- & \textbf{0.104} & \textbf{0.158} \\
		&  &  & 5 & 0.140 & 0.203 \\ \midrule
		Materials Project & \multirow{2}{*}{formation energy} & \multirow{2}{*}{eV / atom} & (hard) 5 & \textbf{0.037} & 0.091 \\
		(N=62k) & & & 5 & 0.039 & \textbf{0.084} \\
		\bottomrule
	\end{tabular}
\end{table}
Table~\ref{tab:results} shows the performance of SchNet models trained on various datasets with and without cutoff.
We observe that the cutoff is beneficial for QM9 as well as the small aspirin training set with N=1,000 reference calculations.
The cutoff function biases the model towards learning from local interactions, which helps with generalization since energy contributions from interactions at larger distances are much harder to disentangle.
On the other hand, the SchNet model with trained on 50,000 aspirin reference calculation benefits from the large chemical environment when not applying a cutoff.
This is because with such a large amount of training data, the model is now also able to infer less local interaction within the molecule.
In case of the materials data, we observe that cosine cutoff and hard cutoff yield comparable results, where the cosine cutoff is slightly preferable since it obtains the lower root mean squared error (RMSE).
Since the RMSE puts more emphasis on larger errors than MAE, this indicates that the cosine cutoff improves generalization.
This may be due to the more local model which is obtained by focusing on smaller distances or by eliminating the discontinuities that are introduced by a hard cutoff.

\begin{figure}[tb]
	\centering
	\includegraphics[width=\textwidth]{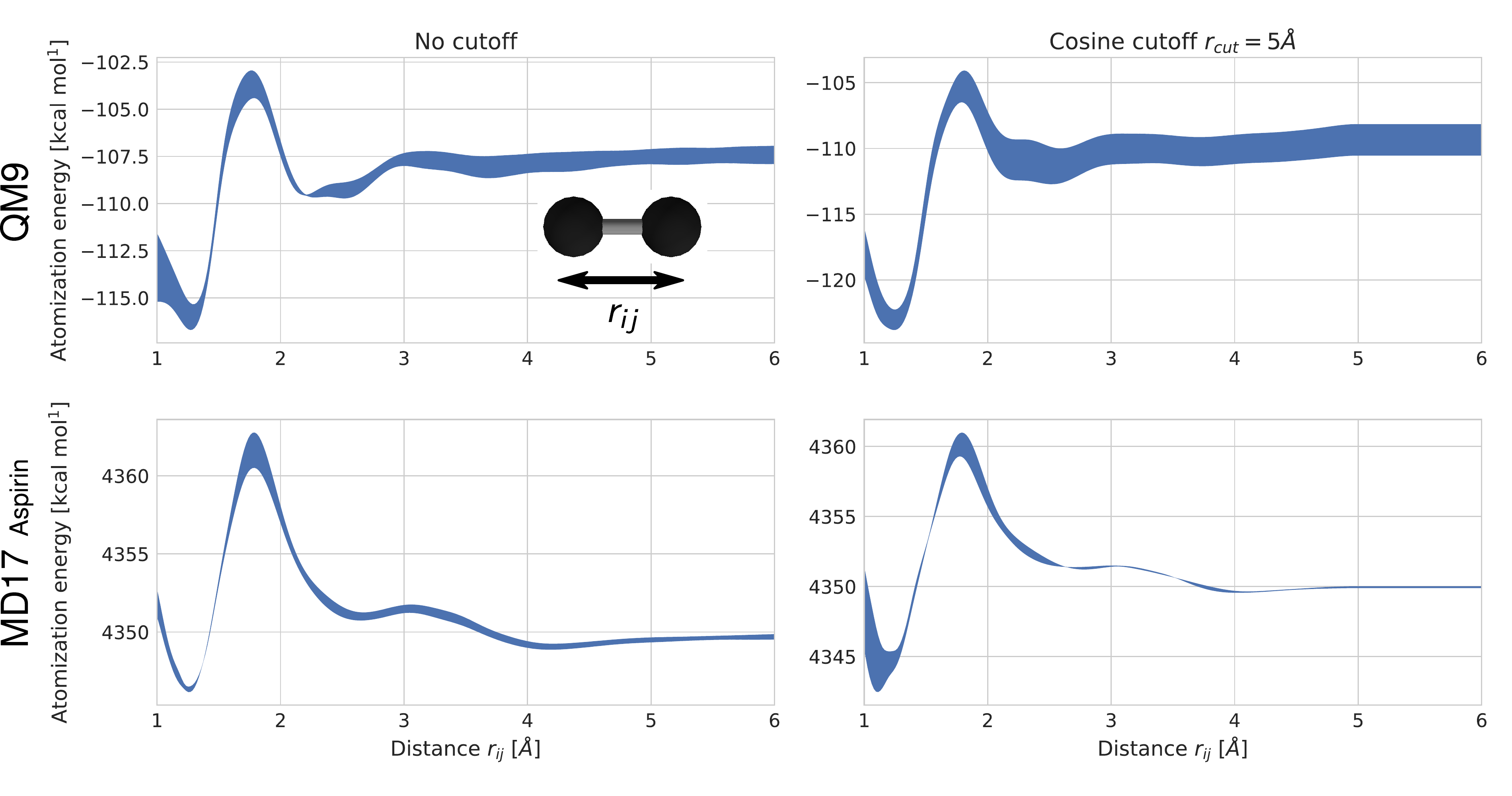}
	\caption{Bond breaking of a carbon dimer as predicted by SchNet trained on QM9 and an aspirin MD trajectory (N=50k). Since the neural networks were not explicitly trained on carbon dimers, the predicted energies are heavily influenced by inferred neighboring atoms. The width of the line represents the deviation of the energy over three models trained on different training splits.}\label{fig:schnet:bondbreaking}
\end{figure}
Fig.~\ref{fig:schnet:bondbreaking} shows the atomization energies of a carbon dimer as predicted by SchNet models trained on QM9 and the aspirin trajectory of the MD17 dataset.
Since the models were trained on saturated molecules, this does not reflect the real energy or atomic forces of the dimer.
The reason is that the energy contribution of carbon interactions in the context of equilibrium molecules or MD trajectories, respectively, include the inferred contributions of other neighboring atoms.
For instance, if we consider two carbon atoms at a distance of about 2.4{\AA} in aspirin, they are likely to be part of the aromatic ring with other carbon at a distance of 1.4{\AA}.
In case of aspirin, we also observe a large offset since the model was not trained on molecules with a varying number of atoms.
If we wanted to eliminate these model biases, we needed to train the neural networks on more diverse datasets, e.g. by explicitly including dimers with large interatomic distances.
While this is necessary to obtain a general model of quantum chemistry, it might even be detrimental for the prediction of a certain subset of molecules using a give amount of training data.

Considering the above, the analysis in Fig.~\ref{fig:schnet:bondbreaking} shows how the neural network attributes interaction energies in the context of the data it was trained on.
We observe that the general shape of the potential is consistent across all four models.
Applying the cosine cutoff leads to constant energy contributions beyond $r_\text{cut}=5${\AA}, however, models without cutoff are nearly constant in this distance regime as well.
SchNet correctly characterizes the carbon bond with its energy minimum between 1.2-1.3{\AA} and rising energy with larger distances.
For the distance regime beyond about 1.8{\AA}, the inferred, larger environment dominates the attribution of interaction energies.

\begin{figure}[tb]
	\centering
	\includegraphics[width=\textwidth]{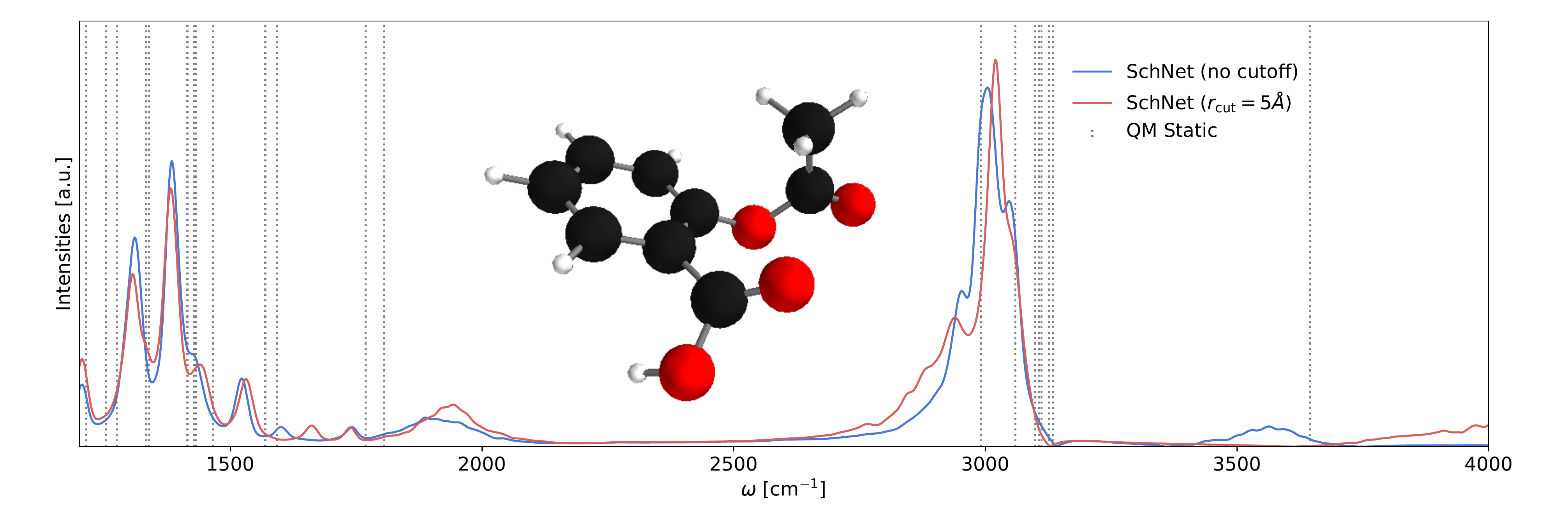}
	\caption{Vibrational spectrum of aspirin as predicted by SchNet without cutoff on 50k reference calculations. The harmonic normal mode vibrations obtained with the electronic structure reference are shown in grey.}\label{fig:aspirin_spectrum}
\end{figure}
Given that the aspirin model trained on the larger dataset benefits from a larger attribution of interaction energies to larger distances, we analyze how the cutoff will affect the vibrational spectrum.
Using the SchNet potentials, we have generated two molecular dynamics trajectories at 300K with a time step of 0.5fs.
Fig.~\ref{fig:aspirin_spectrum} shows the vibrational spectra of the models with and without cosine cutoff.
The most distinct change is the shift of the peak at about 3600 cm$^{-1}$ to the right.
This corresponds to the O-H oscillations, where the cutoff may prevent direct interactions of the hydroxyl group with the carbon ring (see Fig.~\ref{fig:aspirin_spectrum}).

\subsection{Local chemical potentials}
\begin{figure}[tb]
	\centering
	\includegraphics[width=\textwidth]{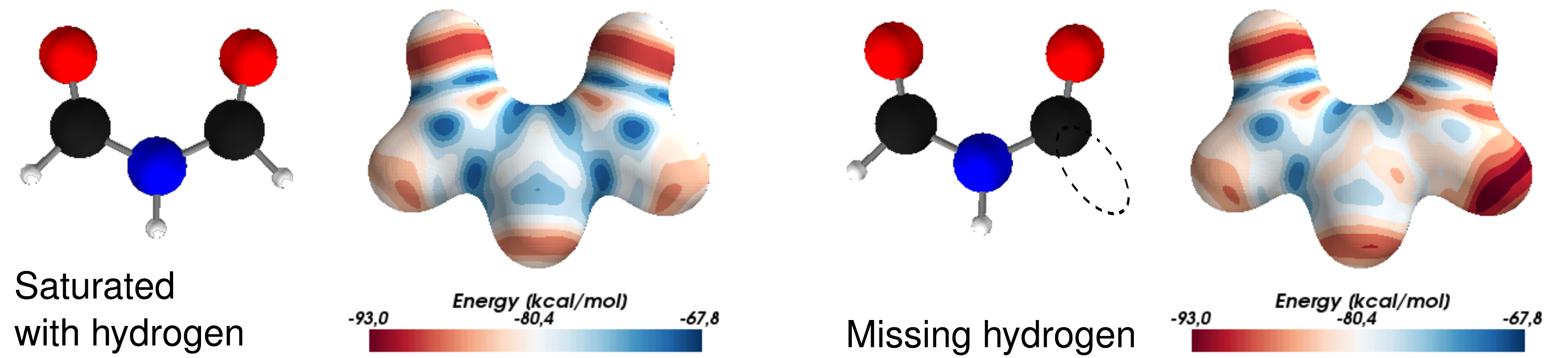}
	\caption{Local chemical potentials of N-Formylformamide generated by SchNet trained on QM9 using a hydrogen probe for the complete molecule (left) and after removing one of the hydrogens (right). The potentials are plotted on the $\sum_i \| \rr - \rr_i \|=3.7${\AA} isosurface of the saturated molecule.}\label{fig:schnet:gbplots}
\end{figure}
In order to further examine the spatial structure of the representation, we observe how SchNet models the influence of a molecule on a probe atom that is moved through space and acts as a test charge.
This can be derived straight-forward from the definition of the continuous-filter convolutional layer in Eq.~\ref{eq:contconvatom}, which is defined for arbitrary positions in space:
\begin{equation}
\x_\text{probe} = (\rho^l * W)(\rr_\text{probe}) = \sum_{j=1}^{n_\text{atoms}} \x^l_j \circ W(\rr_\text{probe}-\rr_j).
\end{equation}
The remaining part of the model is left unchanged as those layers are only applied atom-wise.
Finally, we visualize the predicted probe energy on a smooth isosurface around the molecule~\cite{schutt2017quantum,schutt2018quantum,schutt2018schnet}.

Fig.~\ref{fig:schnet:gbplots} (left) shows this for N-Formylformamide using a hydrogen probe.
According to this, the probe is more likely to bond on the oxygens, as indicated by the lower probe energies.
To further study this interpretation, we remove one of the hydrogens in Fig.~\ref{fig:schnet:gbplots} (right).
In agreement with our analysis, this leads to even lower energies at the position of the missing hydrogen as well as the nearby oxygen due to the nearby unsaturated carbon.

\subsection{Atom embeddings}
\begin{figure}[tb]
	\centering
	\includegraphics[width=0.8\textwidth]{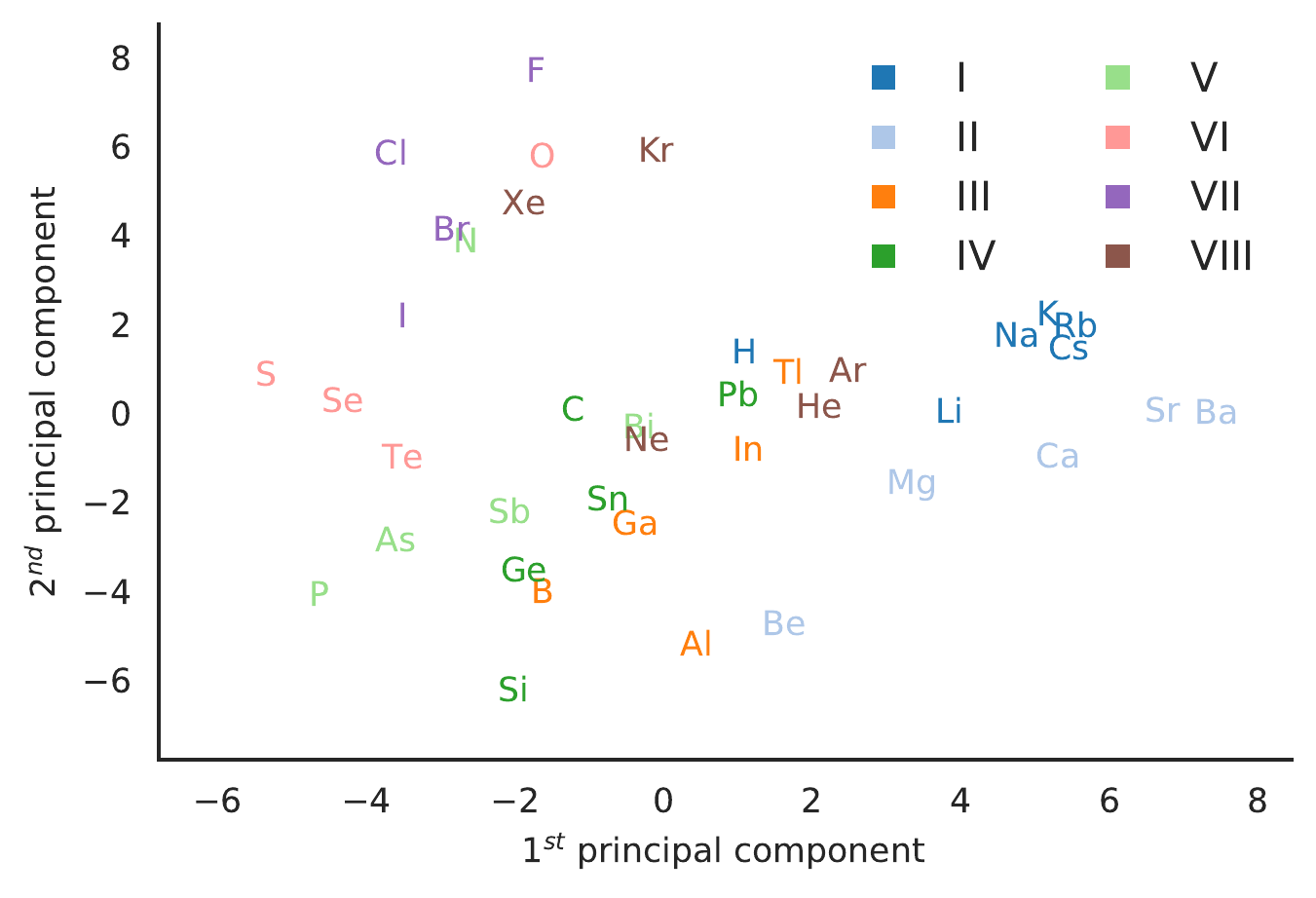}
	\caption{Element embeddings learned by SchNet.}\label{fig:schnet:embeddings}
\end{figure}
Having examined the spatial structure of SchNet representations, we go on to analyze what the model has learned about chemical elements included in the data.
As described above, SchNet encodes atom types using embeddings $\mathbf{A}_{Z} \in \mathbb{R}^F$ that are learned during the training process.
We visualize the two leading principal components of these embeddings to examine whether they resemble chemical intuition.
Since QM9 only contains five atom types (H, C, N, O, F), we perform this analysis on the more diverse Materials Project dataset as it includes 89 atom types ranging across the periodic table.
Fig.~\ref{fig:schnet:embeddings} shows the reduced embeddings of the main group elements of the periodic table.
Atoms belonging to the same group tend to form clusters.

This is especially apparent for main groups 1-7, while group 8 appears to be more scattered.
Beyond that, there are partial orderings of elements according to their period within some of the groups.
We observe a partial order from light to heavier elements in some groups, e.g. in group 1 (left to right: H - Li - Na - [K, Rb, Cs]), group 2 (left to right: Be - Mg - Ca - Sr - Ba) and group 5 (left to right: P-As-Sb-Bi).
In some cases, the first element of the group lies further apart from the rest of the group, e.g. H, Be, C and O.
These results are consistent with those we obtained from previous SchNet models trained on earlier versions of the Materials Project repository~\cite{schutt2018schnet,schutt2018quantum}.

Note that these extracted chemical insights are not imposed by the SchNet architecture, but had to be inferred by the model based on the bulk systems and its energies in the training data.

\section{Conclusions}
We have presented the deep tensor neural network framework and its implementation SchNet, which obtains accurate predictions of chemical properties for molecules and materials.
Representations of chemical environments are learned directly from atom types and position while filter-generating networks allow to incorporate invariance and prior knowledge.

In our analysis, we have found that atomic representations reflect an inferred chemical environment based on the bias of the training data.
The obtained representations are dominated by local interactions and can be further localized using cosine cutoff functions that improve generalization.
However, if a sufficient amount of data is available, interaction energies can be attributed reliably for larger distances which will further improve the accuracy of the model.
Moreover, we have defined local chemical potentials that allow for spatially resolved chemical insights and have shown that the models learn embeddings of chemical elements that show resemblence of the structure of the periodic table.

In conclusion, SchNet presents an end-to-end atomistic neural network that we expect to facilitate further developments towards interpretable deep learning architectures to assist chemistry research.

\begin{acknowledgement}
The authors thank Michael Gastegger for valuable discussions and feedback.
This work was supported by the Federal Ministry of Education and Research (BMBF) for the Berlin Big Data Center BBDC (01IS14013A). 
Additional support was provided by the Institute for Information \& Communications Technology Promotion and funded by the Korean government (MSIT) (No. 2017-0-00451, No. 2017-0-01779). A.T. acknowledges support from the European Research Council (ERC-CoG grant BeStMo).
\end{acknowledgement}

\bibliographystyle{spphys}
\bibliography{references}

\begin{thebibliography}{10}
\providecommand{\url}[1]{{#1}}
\providecommand{\urlprefix}{URL }
\expandafter\ifx\csname urlstyle\endcsname\relax
  \providecommand{\doi}[1]{DOI \discretionary{}{}{}#1}\else
  \providecommand{\doi}{DOI \discretionary{}{}{}\begingroup
  \urlstyle{rm}\Url}\fi

\bibitem{schutt2014represent}
K.T. Sch{\"u}tt, H.~Glawe, F.~Brockherde, A.~Sanna, K.R. M{\"u}ller, E.~Gross,
  Phys. Rev. B \textbf{89}(20), 205118 (2014)

\bibitem{huo2017unified}
H.~Huo, M.~Rupp, arXiv preprint arXiv:1704.06439  (2017)

\bibitem{faber2018alchemical}
F.A. Faber, A.S. Christensen, B.~Huang, O.A. von Lilienfeld, The Journal of
  Chemical Physics \textbf{148}(24), 241717 (2018)

\bibitem{de2016comparing}
S.~De, A.P. Bart{\'o}k, G.~Cs{\'a}nyi, M.~Ceriotti, Physical Chemistry Chemical
  Physics \textbf{18}(20), 13754 (2016)

\bibitem{morawietz2016van}
T.~Morawietz, A.~Singraber, C.~Dellago, J.~Behler, Proceedings of the National
  Academy of Sciences \textbf{113}(30), 8368 (2016)

\bibitem{gastegger2017machine}
M.~Gastegger, J.~Behler, P.~Marquetand, Chem. Sci. \textbf{8}(10), 6924 (2017)

\bibitem{faber2017prediction}
F.A. Faber, L.~Hutchison, B.~Huang, J.~Gilmer, S.S. Schoenholz, G.E. Dahl,
  O.~Vinyals, S.~Kearnes, P.F. Riley, O.A. von Lilienfeld, Journal of chemical
  theory and computation \textbf{13}(11), 5255 (2017)

\bibitem{podryabinkin2017active}
E.V. Podryabinkin, A.V. Shapeev, Computational Materials Science \textbf{140},
  171 (2017)

\bibitem{brockherde2017bypassing}
F.~Brockherde, L.~Voigt, L.~Li, M.E. Tuckerman, K.~Burke, K.R. M{\"u}ller, Nat.
  Commun. \textbf{8}, 872 (2017)

\bibitem{bartok2017machine}
A.P. Bart{\'o}k, S.~De, C.~Poelking, N.~Bernstein, J.R. Kermode, G.~Cs{\'a}nyi,
  M.~Ceriotti, Science advances \textbf{3}(12), e1701816 (2017)

\bibitem{schutt2018schnet}
K.T. Sch{\"u}tt, H.E. Sauceda, P.J. Kindermans, A.~Tkatchenko, K.R. M{\"u}ller,
  The Journal of Chemical Physics \textbf{148}(24), 241722 (2018)

\bibitem{chmiela2018towards}
S.~Chmiela, H.E. Sauceda, K.R. M{\"u}ller, A.~Tkatchenko, arXiv preprint
  arXiv:1802.09238  (2018)

\bibitem{ziletti2018insightful}
A.~Ziletti, D.~Kumar, M.~Scheffler, L.M. Ghiringhelli, Nature communications
  \textbf{9}(1), 2775 (2018)

\bibitem{dragoni2018achieving}
D.~Dragoni, T.D. Daff, G.~Cs{\'a}nyi, N.~Marzari, Physical Review Materials
  \textbf{2}(1), 013808 (2018)

\bibitem{bartok2010gaussian}
A.P. Bart{\'o}k, M.C. Payne, R.~Kondor, G.~Cs{\'a}nyi, Phys. Rev. Lett.
  \textbf{104}(13), 136403 (2010)

\bibitem{rupp2012fast}
M.~Rupp, A.~Tkatchenko, K.R. M{\"u}ller, O.A. Von~Lilienfeld, Phys. Rev. Lett.
  \textbf{108}(5), 058301 (2012)

\bibitem{montavon2012learning}
G.~Montavon, K.~Hansen, S.~Fazli, M.~Rupp, F.~Biegler, A.~Ziehe, A.~Tkatchenko,
  A.V. Lilienfeld, K.R. M\"{u}ller, in \emph{Advances in Neural Information
  Processing Systems 25}, ed. by F.~Pereira, C.J.C. Burges, L.~Bottou, K.Q.
  Weinberger (Curran Associates, Inc., 2012), pp. 440--448

\bibitem{hansen2013assessment}
K.~Hansen, G.~Montavon, F.~Biegler, S.~Fazli, M.~Rupp, M.~Scheffler, O.A.
  Von~Lilienfeld, A.~Tkatchenko, K.R. M\"uller, J. Chem. Theory Comput.
  \textbf{9}(8), 3404 (2013)

\bibitem{Hansen-JCPL}
K.~Hansen, F.~Biegler, R.~Ramakrishnan, W.~Pronobis, O.A. {von Lilienfeld},
  K.R. M{\"u}ller, A.~Tkatchenko, J. Phys. Chem. Lett. \textbf{6}, 2326 (2015)

\bibitem{chmiela2017machine}
S.~Chmiela, A.~Tkatchenko, H.E. Sauceda, I.~Poltavsky, K.T. Sch{\"u}tt, K.R.
  M{\"u}ller, Sci. Adv. \textbf{3}(5), e1603015 (2017)

\bibitem{behler2007generalized}
J.~Behler, M.~Parrinello, Phys. Rev. Lett. \textbf{98}(14), 146401 (2007)

\bibitem{bartok2013representing}
A.P. Bart{\'o}k, R.~Kondor, G.~Cs{\'a}nyi, Phys. Rev. B \textbf{87}(18), 184115
  (2013)

\bibitem{gastegger2018wacsf}
M.~Gastegger, L.~Schwiedrzik, M.~Bittermann, F.~Berzsenyi, P.~Marquetand, J.
  Chem. Phys. \textbf{148}(24), 241709 (2018)

\bibitem{pronobis2018capturing}
W.~Pronobis, K.T. Sch{\"u}tt, A.~Tkatchenko, K.R. M{\"u}ller, The European
  Physical Journal B \textbf{91}(8), 178 (2018)

\bibitem{sifain2018discovering}
A.E. Sifain, N.~Lubbers, B.T. Nebgen, J.S. Smith, A.Y. Lokhov, O.~Isayev, A.E.
  Roitberg, K.~Barros, S.~Tretiak, The journal of physical chemistry letters
  \textbf{9}(16), 4495 (2018)

\bibitem{yao2018tensormol}
K.~Yao, J.E. Herr, D.W. Toth, R.~Mckintyre, J.~Parkhill, Chemical science
  \textbf{9}(8), 2261 (2018)

\bibitem{schutt2018quantum}
K.T. Sch{\"u}tt, M.~Gastegger, A.~Tkatchenko, K.R. M{\"u}ller, arXiv preprint
  arXiv:1806.10349  (2018)

\bibitem{schutt2017quantum}
K.T. Sch{\"u}tt, F.~Arbabzadah, S.~Chmiela, K.R. M{\"u}ller, A.~Tkatchenko,
  Nat. Commun. \textbf{8}, 13890 (2017)

\bibitem{schutt2017schnet}
K.T. Sch\"{u}tt, P.J. Kindermans, H.E. Sauceda, S.~Chmiela, A.~Tkatchenko, K.R.
  M\"{u}ller, in \emph{Advances in Neural Information Processing Systems 30}
  (2017), pp. 992--1002

\bibitem{pukrittayakamee2009simultaneous}
A.~Pukrittayakamee, M.~Malshe, M.~Hagan, L.~Raff, R.~Narulkar, S.~Bukkapatnum,
  R.~Komanduri, J. Chem. Phys. \textbf{130}(13), 134101 (2009)

\bibitem{malshe2009development}
R.~Malshe, M .and~Narulkar, L.M. Raff, M.~Hagan, S.~Bukkapatnam, P.M. Agrawal,
  R.~Komanduri, J. Chem. Phys. \textbf{130}(18), 184102 (2009)

\bibitem{lecun1989backpropagation}
Y.~LeCun, B.~Boser, J.S. Denker, D.~Henderson, R.E. Howard, W.~Hubbard, L.D.
  Jackel, Neural Comput. \textbf{1}(4), 541 (1989)

\bibitem{BrabandereJTG16}
X.~Jia, B.~De~Brabandere, T.~Tuytelaars, L.V. Gool, in \emph{Advances in Neural
  Information Processing Systems 29}, ed. by D.D. Lee, M.~Sugiyama, U.V.
  Luxburg, I.~Guyon, R.~Garnett (2016), pp. 667--675

\bibitem{dugas2001incorporating}
C.~Dugas, Y.~Bengio, F.~B{\'e}lisle, C.~Nadeau, R.~Garcia, in \emph{Advances in
  neural information processing systems} (2001), pp. 472--478

\bibitem{clevert2015fast}
D.A. Clevert, T.~Unterthiner, S.~Hochreiter, arXiv preprint arXiv:1511.07289
  (2015)

\bibitem{chollet2017xception}
F.~Chollet, in \emph{Proceedings of the IEEE Conference on Computer Vision and
  Pattern Recognition} (2017)

\bibitem{szegedy2016rethinking}
C.~Szegedy, V.~Vanhoucke, S.~Ioffe, J.~Shlens, Z.~Wojna, in \emph{Proceedings
  of the IEEE Conference on Computer Vision and Pattern Recognition} (2016),
  pp. 2818--2826

\bibitem{deng2009imagenet}
J.~Deng, W.~Dong, R.~Socher, L.J. Li, K.~Li, L.~Fei-Fei, in \emph{Computer
  Vision and Pattern Recognition, 2009. CVPR 2009. IEEE Conference on} (IEEE,
  2009), pp. 248--255

\bibitem{broomhead1988multivariable}
D.~Broomhead, D.~Lowe, Complex Systems \textbf{2}, 321 (1988)

\bibitem{moody1989fast}
J.~Moody, C.J. Darken, Neural computation \textbf{1}(2), 281 (1989)

\bibitem{ramakrishnan2014quantum}
R.~Ramakrishnan, P.O. Dral, M.~Rupp, O.A. von Lilienfeld, Sci. Data \textbf{1}
  (2014)

\bibitem{loshchilov2016sgdr}
I.~Loshchilov, F.~Hutter, arXiv preprint arXiv:1608.03983  (2016)

\bibitem{kingma2014adam}
D.P. Kingma, J.~Ba, arXiv preprint arXiv:1412.6980  (2014)

\end{thebibliography}

\end{document}